\documentclass[aps,prb,reprint,superscriptaddress,nofootinbib]{revtex4-2}

\usepackage{amsmath,amssymb,bm}
\usepackage{graphicx}
\usepackage[colorlinks=true,citecolor=blue,linkcolor=blue,urlcolor=blue]{hyperref}

\newcommand{\eps}{\varepsilon}
\newcommand{\QC}{Q_{\mathrm{C}}}
\newcommand{\QCz}{Q_{\mathrm{C}0}}
\newcommand{\QIC}{Q_{\mathrm{IC}}}
\newcommand{\DC}{\Delta_{\mathrm{C}}}
\newcommand{\DIC}{\Delta_{\mathrm{IC}}}
\newcommand{\kcdw}{\kappa_{\mathrm{CDW}}}
\newcommand{\keff}{\kappa_{\mathrm{eff}}}
\newcommand{\TIC}{T_{\mathrm{IC}}}

\begin{document}

\title{
  Landau Theory for Commensurate Charge-Density Waves
  Coupled to Uniform Lattice Deformation
}

\author{Keiji Nakatsugawa}
\affiliation{Department of Chemistry \& Biotechnology, School of Engineering, The University of Tokyo, Tokyo 113-8656, Japan}

\author{Toshiyuki Fujii}
\affiliation{Department of Physics, Asahikawa Medical University, Asahikawa 078-8510, Japan}

\author{Satoshi Tanda}
\affiliation{Research Institute for Electronic Science, Hokkaido University, Sapporo, 060-8628, Japan}

\date{\today}

\begin{abstract}
  We formulate a minimal Landau theory for a charge-density wave (CDW) whose commensurability is defined with respect to a deformed lattice. The motivation is provided by recent observations on an isolated single NbS\textsubscript{3} chain, which exhibits a commensurate CDW state accompanied by a $6\%$ shrinkage of the lattice constant. A uniform stretch $a_0\to a_0(1+\eps)$ changes the reciprocal lattice wave number to $G(\eps)=G_0/(1+\eps)$, so that an $N$-fold commensurate CDW has the wave number $\QC(\eps)=G(\eps)/N$, whereas the wave number $\QIC$ favored by the incommensurate instability remains fixed. We propose an amplitude--strain free energy for both $N=3$ and $N=4$, in which the CDW induces a finite uniform strain by relieving the mismatch between $\QC(\eps)$ and $\QIC$. The mismatch is shared between the CDW and the lattice in a proportion set by their stiffness ratio; since the CDW stiffness grows with the CDW amplitude, the lattice takes up an increasing share of the mismatch as the CDW develops.
  Our results suggest a reexamination of lock-in theories and of strain-tuning experiments on density-wave systems.
\end{abstract}

\maketitle

\section{Introduction}

Charge-density wave (CDW) phases in low-dimensional materials are usually described as the result of a competition between an incommensurate instability and a commensurate lock-in imposed by the underlying lattice.
In McMillan's free-energy theory of the CDW phase transition, the CDW order parameter couples to lattice-periodic coefficients, and Umklapp terms stabilize commensurate states when the CDW wave number is commensurate with the reciprocal lattice vector~\cite{McMillan1975,McMillan1976}.
Nakanishi and Shiba further emphasized that incommensurate states near a commensurate lock-in involve higher harmonics, leading to discommensurations~\cite{NakanishiShiba1977,NakanishiShiba1978,Nakanishi1978}.

In all of these theories the lattice enters only as a rigid, externally prescribed periodic background.
The commensurate wave number is therefore a fixed number, and the entire mismatch between the incommensurate and the commensurate wave number must be absorbed by the CDW itself.

These free-energy theories, however, cannot address recent experiments on isolated single-chain NbS$_3$ synthesized within carbon nanotube sheaths~\cite{Tanda2026}.
In the single-chain system a CDW close to $(1/4)b^\ast$ was observed, in contrast to the $(1/3)b^\ast$ CDW known in bulk samples, and the commensurate CDW formation is accompanied by a $6\%$ shrinkage of the lattice constant~\cite{Tanda2026}.
A deformation of this magnitude lies well outside the regime in which the lattice may be regarded as rigid, and it suggests that the CDW wave number and the lattice spacing should be treated as coupled degrees of freedom.

We therefore ask how commensurability should be formulated when the lattice constant itself is allowed to change.
The central idea is simple.
If the lattice is uniformly deformed, the reciprocal lattice wave number changes, and hence the commensurate CDW wave number changes as well.
The wave number favored by the incommensurate instability, on the other hand, is set by the Fermi wave number and is independent of the lattice geometry.
A lattice deformation can therefore change the mismatch between the commensurate and the incommensurate wave numbers, and lock-in may be achieved by deforming the lattice rather than the CDW.
The purpose of this article is to construct a minimal free energy that captures this mechanism.


\section{Free energy}
We consider the free energy
\begin{equation}
  F[\psi,\psi^{*},\eps]=\int dx\; f[\psi,\psi^{*},\eps](x),
  \label{eq:F}
\end{equation}
where $\psi$ is the complex scalar order parameter of the CDW and $\eps$ describes the uniform change of the lattice constant, $a_{0}\to a_{0}(1+\eps)$.
The free energy density is given by
\begin{align}
  f[\psi,\psi^{*},\eps](x)
  ={}& a(x)\alpha(x)^{2}+b(x)\alpha(x)^{3}+c(x)\alpha(x)^{4}
  \nonumber\\
  &+e(x)\bigl|[-i\partial_{x}-\QIC]\psi(x)\bigr|^{2}
  +\frac{\kappa}{2}\eps^{2}.
  \label{eq:f_E}
\end{align}
The charge-density amplitude $\alpha(x)$ is the real part of the order parameter,
\begin{equation}
  \alpha(x)=\operatorname{Re}\psi(x)=\frac{\psi(x)+\psi^{*}(x)}{2},
  \label{eq:alpha}
\end{equation}
and the lattice-periodic coefficients are defined as
\begin{equation}
  a(x)=\sum_{m\in\mathbb{Z}}a_{m}e^{imG(\eps)x},
  \label{eq:ax}
\end{equation}
where reality of the free energy implies $a_{m}=a_{-m}^\ast=|a_{m}|e^{i\theta_{a_m}}$, with analogous expressions for $b(x)$, $c(x)$ and $e(x)$.
Here, we assume that the Fourier amplitudes $a_m$, $b_m$, $c_m$ and $e_m$ are unaffected to leading order in $\eps$, so that the deformation enters only through the phase factor.
This is the Cauchy--Born rule familiar from lattice dynamics~\cite{BornHuang1954}, whose range of validity has been analyzed in detail~\cite{Ericksen2008CauchyBorn,EMing2007CauchyBorn}.
The gradient term proportional to $e(x)$ penalizes deviations of the CDW wave number from $\QIC$, the value favored by the incommensurate instability.
The term proportional to $\eps^{2}$ is the elastic energy of the lattice, where $\kappa>0$ is the stiffness of the lattice.
The form $\eps^2$ is valid for small deformations, $|\eps|\ll1$, and we work consistently to this order throughout.
The free energy density considered by McMillan and by Nakanishi and Shiba is recovered in the limit $\eps\to0$.

An $N$-fold commensurate CDW with respect to the deformed lattice satisfies
\begin{equation}
  N\QC(\eps)=G(\eps)=\frac{G_{0}}{1+\eps},\qquad G_{0}=\frac{2\pi}{a_{0}},
  \label{eq:comm}
\end{equation}
or, equivalently,
\begin{equation}
  \QC(\eps)=\frac{\QCz}{1+\eps},\qquad \QCz=\frac{2\pi}{Na_{0}} .
  \label{eq:QC}
\end{equation}
Thus $\QCz$ is the commensurate wave number of the undeformed lattice and $\QC(\eps)$ is that of the deformed lattice.
The incommensurate wave number, in contrast, is fixed by the electronic structure, $\QIC=2k_\mathrm{F}$ with $k_\mathrm{F}$ the Fermi wave number, and is independent of $\eps$.

For simplicity we also consider the single-mode ansatz
\begin{equation}
  \psi(x)=\phi\,e^{iQx},
  \qquad
  \phi=\Delta e^{i\theta},
  \qquad
  \Delta\ge0,
  \label{eq:single_mode_N}
\end{equation}
so that $\alpha(x)=\Delta\cos(Qx+\theta)$.
Then the free energy density becomes
\begin{align}
  f(x)
  =&
  a(x)\alpha(x)^2
  +
  b(x)\alpha(x)^3
  +
  c(x)\alpha(x)^4
  \nonumber\\
  &+
  e(x)\Delta^2
  \left(
    Q
    -
    \QIC
  \right)^2
  +\frac{\kappa}{2}\eps^2.
  \label{eq:f_single}
\end{align}
In Eq.~\eqref{eq:f_single}, the $m$-th harmonic of the lattice-periodic coefficients, $e^{imG(\eps)x}$, multiplies the $n$-th harmonic of the order parameter, $e^{\pm in(Qx+\theta)}$, and their product is independent of $x$ only if $mG(\eps)=nQ$.
Since $\alpha(x)^{3}$ and $\alpha(x)^{4}$ contain $n=3$ and $n=4$, the lowest-order Umklapp terms are the $b_{1}$ term for $N=3$ and the $c_{1}$ term for $N=4$, which are the commensurabilities relevant to bulk and single-chain NbS$_3$.

\section{Incommensurate state}
In the incommensurate state the CDW wave number takes its favored value $Q=\QIC$, so that the gradient term vanishes and no Umklapp term contributes.
The free energy density of the incommensurate state is therefore
\begin{equation}
  f(x)
  =
  \frac{a_0}{2}\DIC^2
  +
  \frac{3c_0}{8}\DIC^4
  +
  \frac{\kappa}{2}\eps^2.
  \label{eq:f_IC}
\end{equation}
Minimization of the free energy with respect to $\DIC$ and $\eps$ implies
\begin{equation}
  \qquad
  \DIC=\sqrt{-\frac{2a_0}{3c_0}},
  \qquad
  \eps=0,
  \label{eq:Delta_IC}
\end{equation}
for $a_0<0$ and $c_0>0$.
Therefore, the incommensurate CDW does not affect the lattice constant.

\section{Commensurate state}
\subsection{The free energy density}
In the commensurate state the CDW wave number is locked to the deformed lattice, $Q=\QC(\eps)$, and the free energy density is
\begin{align}
  f(x)
  =&
  \frac{a_0}{2}\DC^2
  +
  \frac{|b_{1}|}{4}\DC^3\cos(\theta_{b_1}-3\theta)\delta_{N,3}
  \nonumber
  \\
  &+
  \frac{1}{8}\DC^4\left(3c_0+|c_1|\cos(\theta_{c_1}-4\theta)\delta_{N,4}\right)
  \label{eq:f_C}
  \\
  &+
  f_\eps.
  \nonumber
\end{align}
Here $f_\eps$ collects the geometric-misfit energy of the CDW and the elastic energy of the lattice,
\begin{align}
  f_\eps&=e_0\DC^2
  \left(
    \QC(\eps)
    -
    \QIC
  \right)^2
  +\frac{\kappa}{2}\eps^2
  \nonumber
  \\
  &\approx e_0\DC^2
  \left(
    \QCz(1-\eps)
    -
    \QIC
  \right)^2
  +\frac{\kappa}{2}\eps^2
  \nonumber
  \\
  &=\frac{\kcdw}{2}\left(1-r-\eps\right)^2
  +\frac{\kappa}{2}\eps^2,
  \label{eq:f_eps}
\end{align}
where we defined the stiffness of the CDW as
\begin{align}
  \kcdw\left(\DC\right)=2e_0\DC^2\QCz^2,
  \label{eq:kappa_CDW}
\end{align}
and the dimensionless geometric misfit ratio
\begin{equation}
  r=\frac{\QIC}{\QCz}.
  \label{eq:r}
\end{equation}

Since the global phase $\theta$ is a free variational parameter, the Umklapp terms are minimized by the condition
\begin{equation}
  \theta_{N}-N\theta=(2n+1)\pi,\qquad n\in\mathbb{Z},
  \label{eq:lock}
\end{equation}
with $\theta_N=\theta_{b_1}$ or $\theta_{c_1}$, hence the resulting free energy density is
\begin{align}
  f(x)
  =&
  \frac{a_0}{2}\DC^2
  +
  \frac{3c_0}{8}\DC^4
  -
  \frac{|b_{1}|}{4}\DC^3\delta_{N,3}
  -\frac{|c_1|}{8}\DC^4\delta_{N,4}
  \nonumber
  \\
  &+
  \frac{\kcdw}{2}
  \left(1-r-\eps\right)^2
  +\frac{\kappa}{2}\eps^2.
  \label{eq:f_C_locked}
\end{align}


\subsection{Change in lattice constant}
The equilibrium strain follows from
\begin{equation}
  \frac{\partial f}{\partial\eps}=0,
  \label{eq:stationary}
\end{equation}
which gives
\begin{equation}
  -\kcdw
  \left(1-r-\eps\right)
  +
  \kappa\eps
  =0
  \label{eq:balance}
\end{equation}
and implies
\begin{equation}
  \eps
  =\frac{1-r}{S+1}
  \label{eq:eps}
\end{equation}
where we defined the stiffness ratio
\begin{align}
  S\left(\DC\right)=\frac{\kappa}{\kcdw\left(\DC\right)}.
  \label{eq:S_def}
\end{align}
Equation~\eqref{eq:balance} states that the same restoring force acts on both subsystems, so that the fixed misfit $1-r$ is divided between the lattice, which is deformed by $\eps$, and the CDW, which is deformed by $1-r-\eps$, in the ratio of their stiffnesses,
\begin{equation}
  \frac{\eps}{1-r-\eps}=\frac{\kcdw}{\kappa}.
  \label{eq:ratio}
\end{equation}
As for two Hookean springs in series subjected to a fixed total elongation, the softer subsystem deforms more.
Consistently, substituting Eq.~\eqref{eq:eps} back into Eq.~\eqref{eq:f_eps} collapses the two strain energies into a single term,
\begin{equation}
  f_\eps
  =\frac{\kappa}{2}\frac{(1-r)^{2}}{S+1}
  =\frac{\keff}{2}(1-r)^{2},
  \label{eq:series}
\end{equation}
with the series stiffness
\begin{equation}
  \keff=\frac{\kappa}{1+S}=\frac{\kappa\,\kcdw}{\kappa+\kcdw}.
  \label{eq:kappa_eff}
\end{equation}

\begin{figure}
  \includegraphics[width=\linewidth]{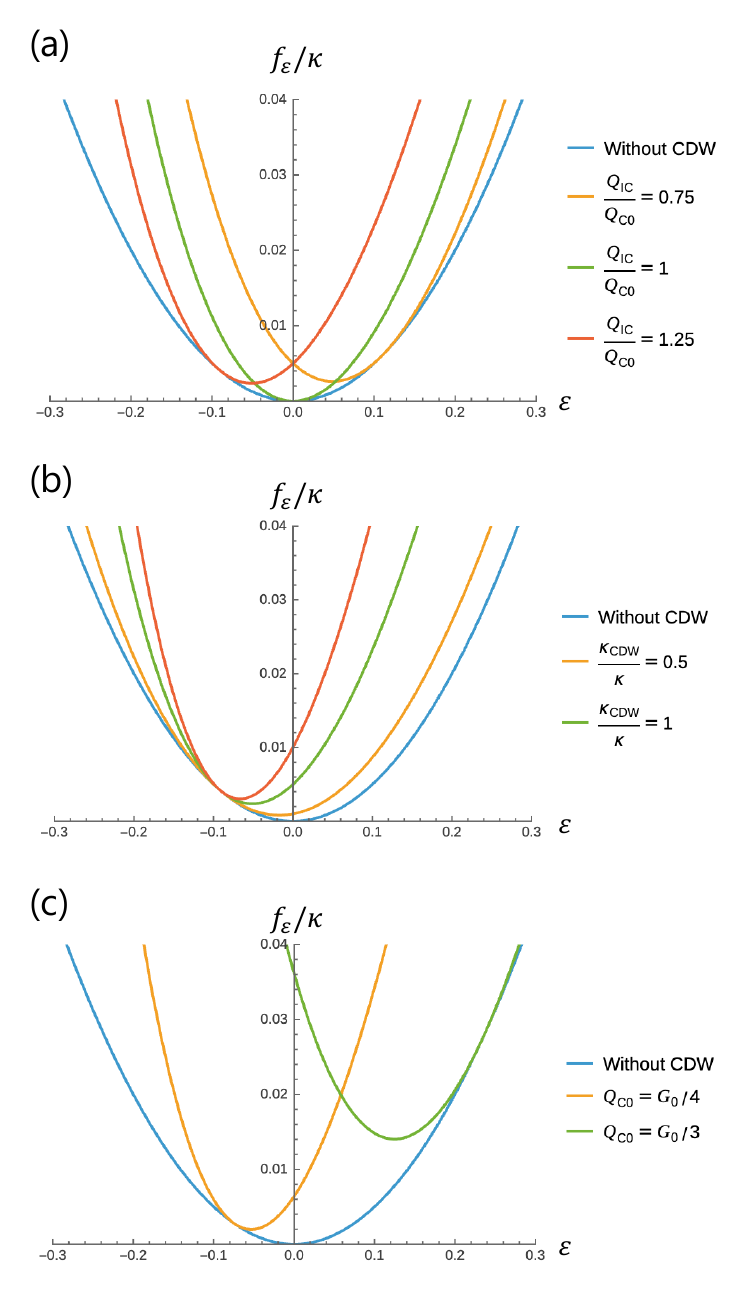}
  \caption{Strain free energy density $f_\eps=\frac{\kcdw}{2}(1-r-\eps)^2+\frac{\kappa}{2}\eps^2$, in units of $\kappa$, as a function of $\eps$.
  (a) The ratio $r=\QIC/\QCz$ determines whether the strain is positive or negative: $r>1$ displaces the minimum to $\eps<0$ and the lattice shrinks, $r<1$ displaces it to $\eps>0$ and the lattice expands, and $r=1$ leaves the minimum at the origin.
  (b) The ratio $S=\kappa/\kcdw$ affects the magnitude of the strain at fixed $r$: the softer the lattice relative to the CDW, the further the minimum is displaced from the origin, up to the bound $\eps=1-r$.
  (c) Dependence on the order of commensurability through $\QCz=G_0/N$, shown for $N=4$ and $N=3$ at fixed $\QIC$; since $r\propto N$, the two commensurabilities correspond to different misfits and hence to different equilibrium strains.
  In all panels the curve labeled ``without CDW'' is the bare elastic energy $\frac{\kappa}{2}\eps^2$.}
  \label{fig:f_varepsilon}
\end{figure}

The change in lattice constant is therefore governed by the two ratios $r$ and $S$.
The sign of $\eps$ is fixed by whether $r$ is larger or smaller than unity, and $\eps$ is nonzero whenever there is a mismatch, $r\ne1$.
The stiffness ratio sets the magnitude: $\eps\to0$ for $S\to\infty$, $\eps\to1-r$ for $S\to0$, and $0<|\eps|<|1-r|$ when $\kappa$ and $\kcdw$ are comparable.
From the definition of the commensurate wave number, Eq.~\eqref{eq:QC}, we also obtain
\begin{equation}
  \frac{\QC(\eps)}{\QCz}=\frac{S+r}{S+1},
  \label{eq:QCratio}
\end{equation}
so that $\QC(\eps)=\QCz$ in the rigid lattice limit $S\to\infty$, as in the previous CDW free-energy theories, whereas $\QC(\eps)=\QIC$ in the soft lattice limit $S\to0$.
These results are summarized in Table~\ref{table:summary} and illustrated in Fig.~\ref{fig:f_varepsilon}.

\begin{table}
  \caption{\label{table:summary}%
  Change of the commensurate wave number $\QC(\eps)$ and of the lattice
  constant $a(\eps)$ as functions of the geometric misfit ratio
  $r=\QIC/\QCz$ and the stiffness ratio $S=\kappa/\kcdw$.}
  \begin{ruledtabular}
  \begin{tabular}{lccc}
    & General $S$ & $S\to\infty$ & $S\to0$ \\
    \colrule
    $\eps$& $\dfrac{1-r}{S+1}$& $0$& $1-r$
    \\[6pt]
    $\QC(\eps)$ & $\dfrac{S+r}{S+1}\,\QCz$&$\QCz$  & $\QIC$
    \\[6pt]
    $a(\eps)$& $\dfrac{S+1}{S+r}\,a_{0}$ & $a_{0}$ & $\dfrac{a_{0}}{r}$ \\
  \end{tabular}
  \end{ruledtabular}
\end{table}

\subsection{Amplitude of the commensurate CDW}
\begin{figure*}
  \includegraphics[width=\linewidth]{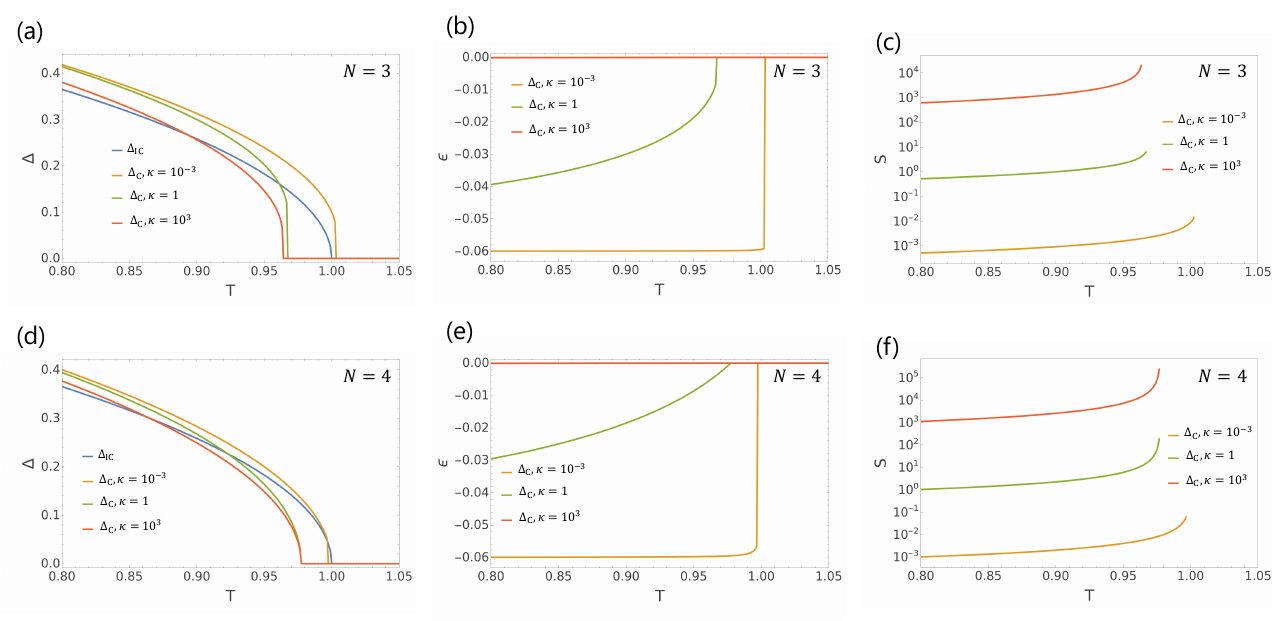}
  \caption{Temperature dependence of the commensurate state for different lattice stiffnesses $\kappa$.
  (a), (d) Comparison of the incommensurate amplitude $\DIC$ and the commensurate amplitude $\DC$ for different values of $\kappa$. In the rigid lattice limit $\kappa\to\infty$ the incommensurate--commensurate phase transition occurs at a temperature $T<\TIC$. In the soft lattice limit $\kappa\to0$ the energy is minimized by deforming the lattice such that ``the lattice is commensurate with the incommensurate CDW wavelength''. For a finite $\kappa$ the incommensurate--commensurate phase transition still occurs, but at a higher temperature.
  (b), (e) The strain $\eps$ increases as $\DC$ develops and saturates at the geometric bound $\eps\to1-r$, equal to $-0.06$ for the parameters used here.
  (c), (f) The stiffness ratio $S$ decreases as the commensurate amplitude develops, implying that the lattice deforms more easily as the temperature decreases.}
  \label{fig:Delta_C}
\end{figure*}

The strain $\eps$ and the stiffness ratio $S$ are both functions of $\DC$, which is obtained by minimizing the free energy.
Substituting the equilibrium strain, Eq.~\eqref{eq:eps}, into Eq.~\eqref{eq:f_C_locked} we obtain
\begin{align}
  f(x)
  =&
  \frac{a_0}{2}\DC^2
  +
  \frac{3c_0}{8}\DC^4
  -
  \frac{|b_{1}|}{4}\DC^3\delta_{N,3}
  -\frac{|c_1|}{8}\DC^4\delta_{N,4}
  \nonumber
  \\
  &+
  \frac{\kappa}{2}\frac{(1-r)^2}{S\left(\DC\right)+1}
  .
  \label{eq:f_C_amp}
\end{align}
To discuss the temperature dependence, we take the conventional Landau form $a_0=a_T(T-T_0)$, while the remaining coefficients are treated as temperature independent.
Equation~\eqref{eq:Delta_IC} implies that $T_0$ is the incommensurate transition temperature $\TIC$.
Minimizing Eq.~\eqref{eq:f_C_amp} with respect to $\DC$, we find that the equilibrium amplitude satisfies
\begin{align}
  0
  =&
  a_0
  +
  \frac{3c_0}{2}\DC^2
  -
  \frac{3|b_{1}|}{4}\DC\delta_{N,3}
  -\frac{|c_1|}{2}\DC^2\delta_{N,4}\nonumber
  \\
  &+
  2e_0\QCz^2(1-r)^2\left[\frac{\kappa}{\kappa+2e_0\DC^2\QCz^2}\right]^2 ,
  \label{eq:Delta_C_eqn}
\end{align}
which we solve numerically; the result is shown in Fig.~\ref{fig:Delta_C}.

In the rigid-lattice limit $\kappa\to\infty$, the conventional rigid-lattice result is recovered.
As the lattice becomes softer, the last factor in Eq.~\eqref{eq:Delta_C_eqn} is reduced because the lattice deformation relaxes the geometric mismatch between $\QC$ and $\QIC$.
Consequently, the commensurate CDW is stabilized and its transition temperature increases toward $\TIC$.
In the soft-lattice limit $\kappa\to0$, the geometric-mismatch contribution to Eq.~\eqref{eq:Delta_C_eqn} vanishes for a finite CDW amplitude: the mismatch is accommodated by lattice deformation rather than by deformation of the CDW wave number.



\section{Discussion}

The present theory proposes a new mechanism of commensurate lock-in and provides a continuous interpolation between two limiting mechanisms.
In the conventional rigid-lattice limit the lattice cannot deform, so that the CDW wave number must change away from its favored wave number $\QIC$ in order to become commensurate.
In the opposite soft-lattice limit the CDW wave number does not change at all; instead the lattice deforms until its commensurate wave number coincides with $\QIC$.
For a finite lattice stiffness both subsystems accommodate the geometric misfit, in the proportion given by Eq.~\eqref{eq:ratio}.

An important feature of the present theory is that the stiffness ratio $S$ of Eq.~\eqref{eq:S_def} depends on the CDW amplitude, $S\propto\DC^{-2}$.
As $\DC$ develops upon cooling, $\kcdw$ increases and $S$ decreases; consequently lattice deformation becomes energetically more favorable relative to deformation of the CDW.
This produces a feedback in which the development of the commensurate CDW enhances the lattice deformation, which in turn reduces the geometric-misfit energy and further stabilizes the CDW.

For $r>1$ the favored incommensurate wave number is larger than the commensurate wave number of the undeformed lattice, and Eq.~\eqref{eq:eps} gives $\eps<0$: the lattice shrinks.
Conversely, $r<1$ produces an expansion.
In the soft-lattice limit $\eps=1-r$, so that a lattice shrinkage of approximately $6\%$ corresponds to $r\approx1.06$.
Thus the lattice contraction observed in an isolated NbS$_3$ chain can be understood as a direct consequence of the geometric mismatch between the incommensurate wave number and the commensurate wave number of the undeformed lattice.

Since $\QCz=G_{0}/N$, the misfit ratio $r=\QIC/\QCz$ is proportional to $N$, and the same $\QIC$ therefore implies different misfits for different commensurabilities.
With $r\approx1.06$ for $N=4$ one obtains $r\approx0.79$ for $N=3$, corresponding to an expansion of order $20\%$, which is far more costly.
Within the present theory the $N=4$ lock-in of the single chain is thus favored over $N=3$ precisely because it requires the smaller lattice deformation, which offers a natural account of the difference between the single-chain and bulk commensurabilities.

More broadly, the present results indicate that the lattice cannot be treated as a rigid background in the physics of commensurate lock-in: the geometric mismatch is itself a relaxable degree of freedom.
Existing lock-in theories, and experiments in which strain is applied to density-wave systems, treat this mismatch as an externally fixed parameter, and both deserve reexamination from this point of view.
Since $\kappa$ measures the stiffness of the entire structure that carries the chain, the soft-lattice regime should be most accessible in low-dimensional and suspended systems, where the lattice is least clamped.
Finally, we note that the single-mode ansatz~\eqref{eq:single_mode_N} excludes discommensurations; allowing a nonuniform phase, and with it a nonuniform strain field, is a natural extension of the present theory.

\end{document}